# A proposal for investigating three-body forces in Aharonov-Bohm systems

Y.Ben-Aryeh

Physics Department , Technion-Israel Institute of Technology

Haifa, Israel

Email:phr65yb@physics.technion.ac.il

Although there is no force on the electron in Aharonov-Bohm solenoid effect, the electron exerts a force on the solenoid related to the inequality of action and reaction forces of two subsystems in three-system configuration. The AB phase which is equal to $-\frac{q}{c\hbar}\oint \vec{A}\cdot d\vec{x}$ is related to the force exerted by the electron on the solenoid which is given as $\vec{F} = -\frac{d}{dt}\left(\frac{q}{c}\vec{A}\right)$. The momentum changes of the mechanical oscillator are equal in magnitude and opposite in sign to the changes in the momentum of the em fields. It is proposed to investigate momentum changes of "micro" bodies producing magnetic fields in AB systems which will clarify the nature of these effects. The problem of magnetic fields shielded from the electron wave packet is also discussed.

## 1. Three-body forces in Aharonov-Bohm solenoid effect

The force that the electron exerts on the solenoid in the simple Aharonov-Bohm system [1] is given by [2,3]

$$\vec{F} = \left(\frac{q}{c^2}\right) \int \vec{j}(\vec{r}\,') \times \left(\vec{v}\times \vec{\nabla}_r\right) |\vec{r}-\vec{r}\,'|^{-1} \, d^3 r' \quad . \tag{1}$$

Here $\vec{r}$ and $\vec{r}'$ are the coordinates of the electron and the solenoid current density $\vec{j}(\vec{r}')$, respectively, and where the integration is taken over all the currents of the solenoid. $q\vec{\nabla}_r |\vec{r}-\vec{r}'|^{-1}$ is the electric field produced by the electron at the current element $\vec{j}(\vec{r}')$ and where $\frac{q}{c}\vec{v}\times\vec{\nabla}_r |\vec{r}-\vec{r}'|^{-1}$ is the corresponding magnetic field. The vector potential at electron position due to solenoid currents is given by

$$\vec{A}(\vec{r}) = \frac{1}{c}\int \vec{j}(\vec{r}') |\vec{r}-\vec{r}'|^{-1} d^3r \quad , \tag{2}$$

Substituting Eq.(2) into Eq. (1) we get

$$\vec{F} = -\left(\frac{q}{c}\right)(\vec{v}\times\vec{\nabla})\times\vec{A}(\vec{r})$$
$$= -\frac{q}{c}[(\vec{v}\cdot\vec{\nabla})\vec{A}(\vec{r}) + \vec{v}\times(\vec{\nabla}\times\vec{A}) - \vec{v}(\vec{\nabla}\cdot\vec{A})] \quad , \tag{3}$$

For stationary currents we assume $\vec{\nabla}\cdot\vec{j} = 0$ leading by Eq. (2) to $\vec{\nabla}\cdot\vec{A} = 0$. The force on the electron represented by $\left(\frac{q}{c}\right)\vec{v}\times(\vec{\nabla}\times\vec{A})$ vanishes, as $(\vec{\nabla}\times\vec{A}) = \vec{B}$ is equal to zero outside the solenoid. We get therefore that the force exerted by the electron on the solenoid is given by :

$$\vec{F} = -\frac{d}{dt}(\frac{q}{c}\vec{A}) \tag{3}$$

The crucial point in this analysis is that the force on the solenoid is equal in magnitude and opposite in sign to the rate of change of momentum of the em field. It has been shown in a previous work [2] that the integral of the change of em momentum over trajectory of the electron, which is equal to the integral of the energy of interference over the corresponding time , gives the AB phase shift.

## 2. Aharonov-Bohm effect with magnetic fields shielded from the electron wave-packet

The basic idea obtained from the above short analysis is that the phase change obtained by the electron in AB effect is intrinsicly coupled with three-body forces in which the mechanical solenoid obtains momentum changes equal in magnitude and opposite in signs with the momentum changes of the em fields. Evidence for AB effect is obtained, however, with magnetic fields shielded from the electron wave packet [4]. The problem arises therefore, how much this shielding is effective in preventing the electric fields of the electron from penetrating into the solenoid and also acting on the currents of the solenoid. For a complete shielding of the em fields from the solenoid the copper layer surrounding the solenoid [4] should be "grounded". Assuming that there are experiments in which a perfect shielding does not affect the AB phase, then our previous calculations for energy of interference are not relevant and the force on the solenoid leads to the inequality $|\vec{F}| < |-\frac{d}{dt}\left(\frac{q}{c}\vec{A}\right)|$. However, even in such cases the electron can exert forces on the currents induced in the copper layer. Therefore even in more general cases the assumption that momentum changes in mechanical parts of the system are equal to the momentum changes off the em fields remains valid. It is quite difficult to investigate the momentum changes in the em fields, but with the new developments in nano-technolgy it might be possible to investigate the force which the electron exerts on the currents producing the magnetic fields or on those induced in the copper layer shielding the magnetic fields.

## 3. Comparison between momentum transfers in AB effect and other optical systems

The problems related to em momentum transfers in AB systems is similar (although it involves a basic different effect) to those explained in electromagnetic Mach-Zehnder interferometers or double slit or "Welcher Weg" experiments [5,6]. A full analysis of such systems should include momentum transfers to the macroscopic entangled parts of the system (beam splitter and mirrors in Mach-Zehnder interferometer and in "Welcher Weg" experiments or macroscopic double-slit wall in 2-slit experiments). It is quite interesting to notice that in Welcher-Weg (WW) measurements, momentum transfers effects should be taken into account. Although the WW apparatus that "determines" which-way might not include any momentum transfer, there is a momentum transfer to the macroscopic entangled parts of the system. Due to the large mass of the macroscopic bodies such momentum transfers are not observed. However, with the development of "nano-technology" one might expect that such momentum transfers can be observed with a "micro"–Mach Zehnder interferometer, a "micro"-wall in double slit experiments, or "micro" Welcher-Weg experiments.

In analogous way (but for a different system) there is no force on the electron in AB effect but it affects momentum transfers between the mechanical parts of the system and the em fields. Although it might be very difficult to observe em momentum transfers expected in AB effects, it might be possible to observe the momentum changes of "micro" bodies producing the magnetic fields. Such experiments with nano-technology might be very difficult to achieve. However, due to the theoretical importance of AB effects it is suggested to try doing such experiments which will clarify the nature of these effects.

4, Conclusion

*Three-bodies forces* occur in AB solenoid effect. Although no force is exerted on the electron it produces momentum transfers between the mechanical solenoid and the em fields. It is suggested to investigate such forces by using "micro" AB systems which might be produced by nano-technology.